\documentclass{cimento}
\usepackage{graphicx}  
\usepackage{amsmath}  
\usepackage{amssymb}  
\usepackage{color}
\definecolor{grey}{rgb}{.7,.7,.7}
\usepackage{braket}
\title{Nanoscale sensing and quantum coherence}
\author{Friedemann~Reinhard\from{ins:tum}}
\instlist{\inst{ins:tum} Technische Universit\"at M\"unchen, Walter Schottky Institut and Physik-Department}

\begin{document}

\maketitle

\begin{abstract}

Small solid state qubits, most prominently single spins in solids, can be remarkable sensors for various physical quantities ranging from magnetic fields to temperature. They package the performance of their bulk semiconductor counterparts into a nanoscale device, sometimes as small as a single atom. This review is a minimalist introduction into this concept. It gives a brief summary of quantum coherence, Ramsey spectroscopy and a derivation of the "standard quantum limit" of the sensitivity that a single-qubit sensor can reach. It goes on to discuss the surprising improvement that dynamical decoupling has brought about and concludes with an outlook to the major frontiers of the field. 
\end{abstract}
The history of quantum mechanics is a history of repeated underestimation. When it emerged in the early 20th century, its strange consequences were discovered only in theory, and illustrated by a series of speculative "Gedankenexperiment"s. Observing them in reality - on single photons, single atoms and single spins - seemed like an impossible mission. \\
When it became reality half a century later, the experimental effort of these low-energy experiments rivaled their early counterparts in high-energy physics. Operating a set of even few qubits involved a lab densely packed with electronics, optics and vacuum setups, and a team of skilled students that would spend their days aligning dye lasers and chasing leaks - an effort comparable to a lab-scale accelerator in the 1930s. Skepticism remained whether these setups would ever make the way out of the lab into the real world. \\
Today, this is a realistic prospect, which is largely owing to some decisive progress of the 2000s and 2010s: the discovery of several qubits that can be implemented in the solid state, in real devices interfaced with the classical world, partly even at room temperature and ambient conditions. With this tool at hand, the attention of the scientific world is increasingly shifting to a very different question: "what to do with it?" rather than "how to do it?". \par
The situation is remarkably similar to the early days of the laser, which was admired as "a solution looking for a problem". This problem, at least a major one, turned out to be sensing. The perfectly monochromatic light of a laser lent itself to measurements of tiny displacements in interferometers, precise spectroscopy of molecules and, since it enabled diffraction-limited focusing, readout of compact discs. It is a tempting idea that perfectly monochromatic matter - which qubits essentially are - will find similar applications. Their narrow spectral lines could respond to tiny shifts imprinted by external fields, making them attractive sensors for various quantities. In those applications where size and effort do not matter, this development has already taken place. Atomic clocks - which can be regarded a quantum sensor for frequency and phase - have become the state of the art for timekeeping and have transformed our life by enabling satellite navigation. Finding similar applications for the younger solid state qubits remains an exciting challenge. One obvious window of opportunity is sensing at the nanoscale. Solid state qubits are the smallest sensors that have ever been conceived - as small as a single atom. Since such a tool has not existed until a decade ago, the mere smallness of nanoscale sensors will open unchartered territory for applications, even in the case that their sensitivity cannot be pushed beyond competing large-scale technology. 

\section{Single molecules and spins as scanning probes}
\begin{figure}
\includegraphics{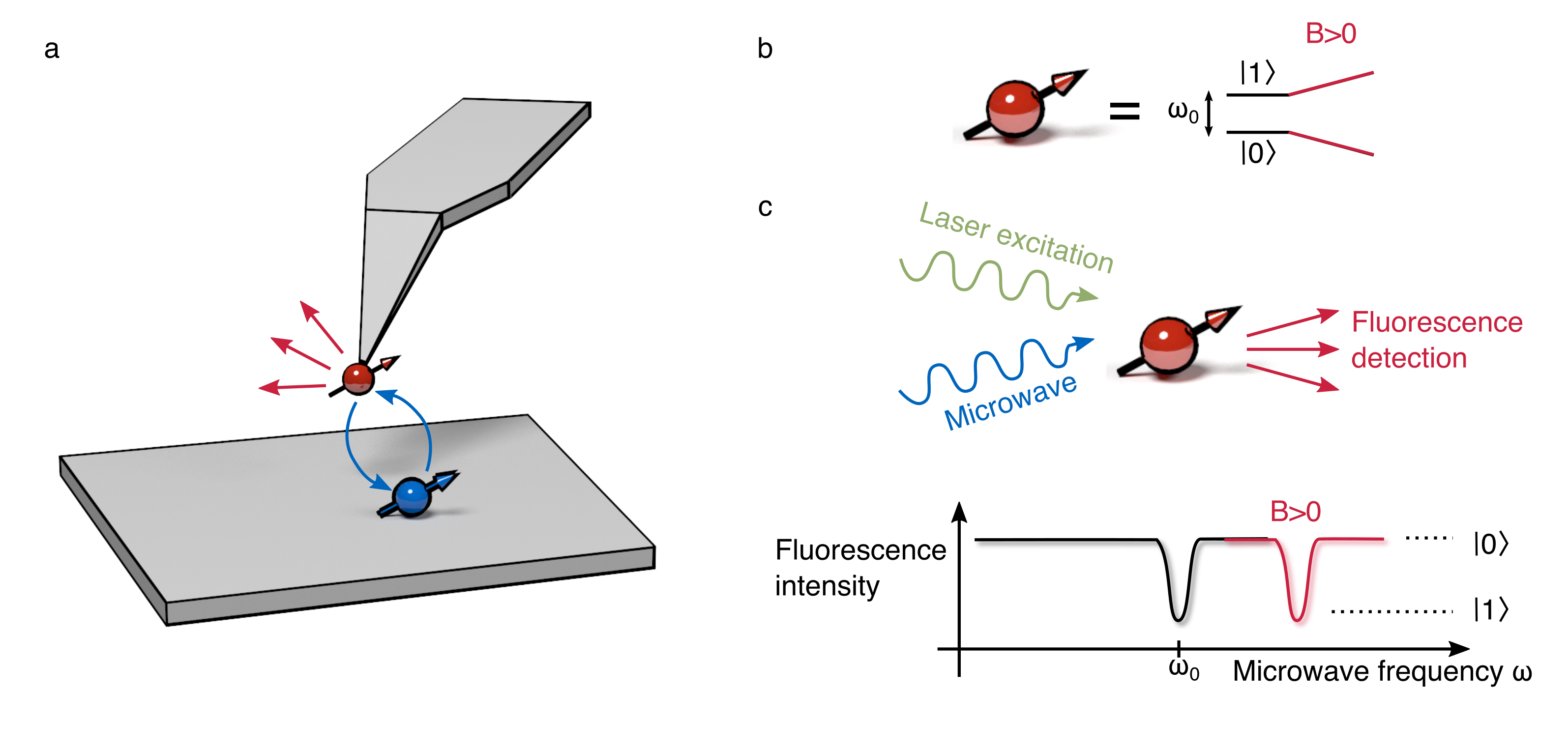}     
\caption{Single spins as scanning probes. a) an optically readable electron spin at the tip of a scanning probe microscope provides a sensor to image single electron and nuclear spins in samples, as well as magnetic fields. b+c) magnetic field sensing by resonant spectroscopy. b) the energy levels of an electron spin are shifted in a magnetic field by the Zeeman effect. c) measurement of the Zeeman shift by optically detected magnetic resonance (ODMR) spectroscopy. Laser illumination pumps the spin into a specific state $\ket 0$. A tuneable microwave switches the spin into another state $\ket 1$ with less fluorescence intensity if it is resonant with the Zeeman-shifted spin transition. }
\end{figure}
When scanning probe microscopy was invented in the 1980s, it quickly became clear that it would become a generic tool, not limited to tunneling microscopy on conductive surfaces. This insight sparked the development of the atomic force microscope~\cite{binnig86}, and subsequent proposals to use it as a novel detector for magnetic fields~\cite{saenz87, martin87} and magnetic resonance microscopy on small ensembles of spins \cite{zuger93}. At roughly the same time, optical detection of single fluorescent molecules became widely available. This catalyzed the development of numerous novel techniques in biology, ranging from fast DNA sequencing~\cite{shendure08} to superresolution microscopy~\cite{hell94,betzig06,rust06}, and led to a strong interest in nanoscale optics. Scanning near-field microscopy was developed, a technique to image optical properties with 10nm-scale resolution by a subwavelength aperture formed by the tip of a pulled glas fiber~\cite{betzig93}. Using a single fluorescent molecule or color center as a scanning probe in this technique became both a realistic vision and an attractive goal. It would push near-field microscopy to its ultimate resolution limit, enabling the study of energy transfer between single molecules, single plasmons and other nano-emitters. This technique has been proposed~\cite{sekatskii96} and demonstrated~\cite{michaelis00}, but failed to find widespread adoption because molecules would photobleach too fast to enable reliable imaging and color centers would reside too deep in their host crystal for photonic coupling.  
\\
Some visionaries were not scared away by these problems and proposed an even more ambitious extension: to use a fluorophore with a stable electron spin, whose state could be optically "read out" by fluorescence~\cite{chernobrod05}. Some fluorophores of this kind were known to exist, most prominently organic molecules with a metastable triplet state. Optically detected magnetic resonance had already been demonstrated in these systems on the level of single stationary molecules in pioneering work in the 1990s~\cite{wrachtrup93, kohler93}. If successful, a scanning probe microscope with such a sensor promised to transform near-field microscopy into a magnetic resonance microscopy technique. Spins in the sample coupling to the sensor spin would cause spin flips that could be read out by fluorescence. The whole arsenal of multidimensional magnetic resonance could be used to infer the location of spins in the sample, potentially with atomic resolution. On a lower technical level, shifts of the sensor spin resonance could be used to measure and image electric fields (via a Stark shift) and magnetic fields (via a Zeeman shift). 
\par
It was around 2008 that this idea suddenly became a realistic prospect, when several groups put forward proposals and demonstrations indicating that NV (Nitrogen-Vacancy) defects in nanodiamonds could provide a photostable fluorescent nanoprobe with an optically readable electron spin~\cite{taylor08, degen08, balasubramanian08, maze08}. These properties had already been established in the 80s~\cite{oort88} and, on the level of single spins, in the 90s~\cite{gruber97}. We will not present them in detail here, since this review will focus on spin sensing and is agnostic with respect to the exact spin employed an the way its readout is implemented. Suffice it to say that spin readout of the NV center is possible, because its fluorescence is slightly spin dependent (Fig. 1b). It is roughly 30\% more intense if the center is prepared in its lowest spin state $\ket 0$, because fluorescence competes with a strong non-radiative pathway when the center is prepared in a higher spin state $\ket 1$. Due to some lucky incidents, this nonradiative path has a preference to decay into the lower spin state $\ket 0$, so that optical excitation not only reads out the spin, but also initializes it into a known state ($\ket 0$). Finally, the energies of these spins states are shifted by magnetic fields, experiencing a field-dependent Zeeman-shift $\Delta \omega = \gamma B$, where $\gamma/2\pi\approx 30$ MHz/mT$=3$ MHz/G denotes the gyromagnetic ratio. Similar shifts exist for other physical quantities, but we will, without loss of generality, focus on magnetic field sensing in the following.\par
These properties are the basis for the simplest sensing experiment - measuring magnetic fields by spectroscopy of the Zeeman shift (Fig. 1c). Here, the sensor spin state is continuously probed by a continous-wave laser and excited by a microwave at a tuneable frequency $\omega$. For most frequencies, the microwave will not be resonant with the spin transition, so that the laser pumps the spin state into the bright state $\ket 0$, causing a high level of fluorescence. When the microwave is tuned into resonance (so that $\omega = \omega_0 + \gamma B$), it will excite the center into the dark state $\ket 1$, which becomes visible as a dip in fluorescence. By this combined microwave-optical spectroscopy, the Zeeman shift can be measured with reasonable precision, from which the magnetic field at the center can be infered. These experiments reach a typical precision in the $\mu T$ range, roughly 1$\%$ of the earth's magnetic field. While this is inferior to most existing semiconductor sensors, it is measured by the smallest sensor that can possibly be conceived, consisting of little more than a single atom. \par
Single NV centers can be attached to the tip of a scanning probe microscope, either by packaging them in nanodiamonds that can be glued to a commercial tip~\cite{balasubramanian08}, or by sculpting all-diamond AFM tips with a single center embedded at their apex~\cite{maletinsky12}. Repeating the Zeeman spectroscopy experiment at every pixel of a scan, magnetic fields can be imaged with higher resolution than the 10nm-scale state of the art achieved by magnetic force microscopy. As an additional benefit, the stray field of the NV spin does not exert backaction on soft-magnetic samples. These advantages have made spin-based scanning probe magnetometry a fertile field of research, and highlight results include images of moving domain walls~\cite{tetienne14, tetienne15}, antiferromagnets~\cite{gross17}, Skyrmions~\cite{dovzhenko18, yu18}, hard disk write heads~\cite{jakobi17} and superconducting vortices~\cite{thiel15, pelliccione15}. 

\section{Sensing by quantum coherence - reaching the fundamental limit of sensitivity}
\subsection{Quantum coherence as a sensor}
The continous-wave protocol of Fig. 1 has a number of technical problems. Laser and microwave power need to be carefully tuned to ensure that the microwave excitation performs exactly one full spin-flip between two fluorescence events. Furthermore, the spectral resolution depends on microwave power, since a high-power pulse can excite spins even if its frequency is slightly off-resonant. \\
It is for these reasons that most experiments today are performed using pulsed protocols, inspired by atomic clocks \cite{ramsey50} and quantum logic. The key idea, presented in Fig. 2, is to employ the quantum-mechanical phase $\phi$ of a coherent superposition $(\ket 0 + e^{i\phi}\ket 1 )/ \sqrt 2$ as a sensor. 
\begin{figure}
\label{fig:bloch_sphere}
\includegraphics{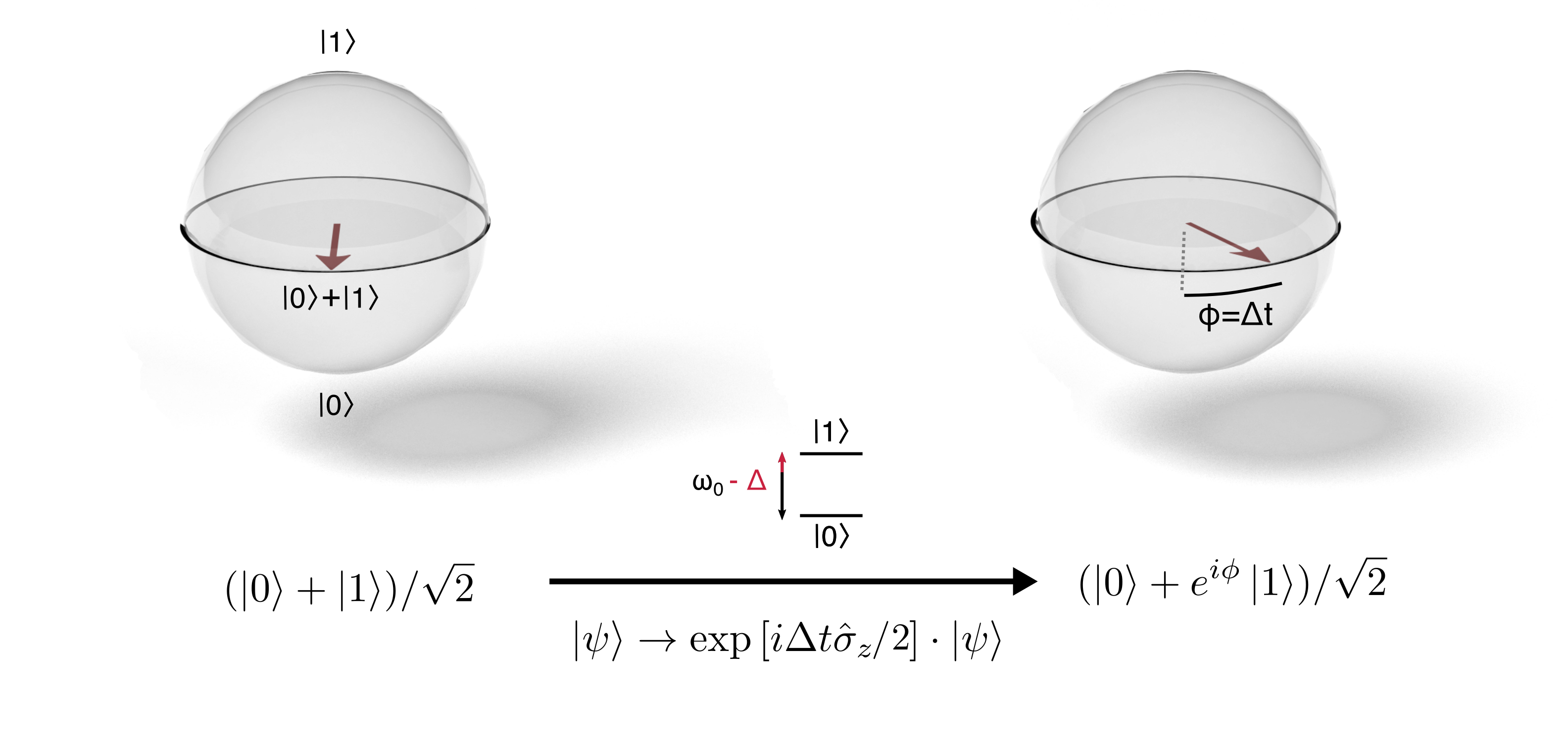}     
\caption{Quantum phase as a sensor for level shifts. The phase $\phi$ of a coherent superposition $(\ket 0 + e^{i\phi}\ket 1) $ grows as $\phi=\Delta t$ under a spectral shift $\Delta$. }
\end{figure}
To see why this phase is sensitive to a small spectral shift $\Delta=\gamma B$, we consider a qubit evolving under the Hamiltonian
$$\hat H = \frac\hbar 2 (\omega_0 - \Delta) \hat \sigma_z \stackrel{\text{rot. frame}}{=} -\frac \hbar 2 \Delta \hat \sigma_z.$$
Here we have applied a transformation into a rotating frame, essentially a renormalization of the qubit energy, to remove the static energy $\omega_0$. The time evolution under this Hamiltonian is
\begin{eqnarray}
\label{eq:free_evolution}
\ket {\psi(t)} &=& \exp\left[-\frac {i  t}\hbar\hat H\right] \ket{\psi(t=0)} \\ 
&=& \exp\left[i \frac{\Delta  t}{2}\hat \sigma_z\right] \ket{\psi(t=0)}\\ &\stackrel{\ket{\psi(t=0)} = (\ket 0 + \ket 1)/\sqrt 2}{=}& (\ket 0 + e^{i\Delta t}\ket 1)/\sqrt 2.
\end{eqnarray}
up to an insignificant global phase $e^{-i\Delta t/2}$. For a nonzero detuning $\Delta$, the phase $\phi$ grows over time. Since this evolution occurs in absence of any manipulation and readout pulses, it is insensitive to experimental fluctuations in laser and microwave power.\par
It is instructive to consider this evolution on the "Bloch sphere" (Fig. \ref{fig:bloch_sphere}). This sphere is a map from $SU(2)$ (spin states) to the unit sphere in $\mathbb{R}^3$. $\ket 0$ and $\ket 1$ are mapped to its south and nord poles respectively. Coherent superpositions $(\ket 0 + e^{i\phi} \ket 1)/\sqrt 2$ reside on the equator, with $\phi$ translating into their geographical longitude. In this picture the time evolution of eq. ($\ref{eq:free_evolution}$) is a rotation of the spin along the equator. Since this rotation revolves with angular velocity $\Delta$, it is very much reminiscent of the movement of the needle in a classical instrument. 
\subsection{Creating and reading out quantum coherence - the Ramsey protocol}
In contrast to a classical needle, the quantum-mechanical phase $\phi$ is invisible. Therefore, the protocol of Fig. 2 needs to be supplemented by two additional steps, one to prepare a coherent superposition and one to convert the phase into a measurable quantity such as the spin state itself (Fig. \ref{fig:ramsey}). Both steps are implemented by microwave pulses, specifically "$\pi/2$" pulses of carefully tuned length.  \par
\begin{figure}
\includegraphics{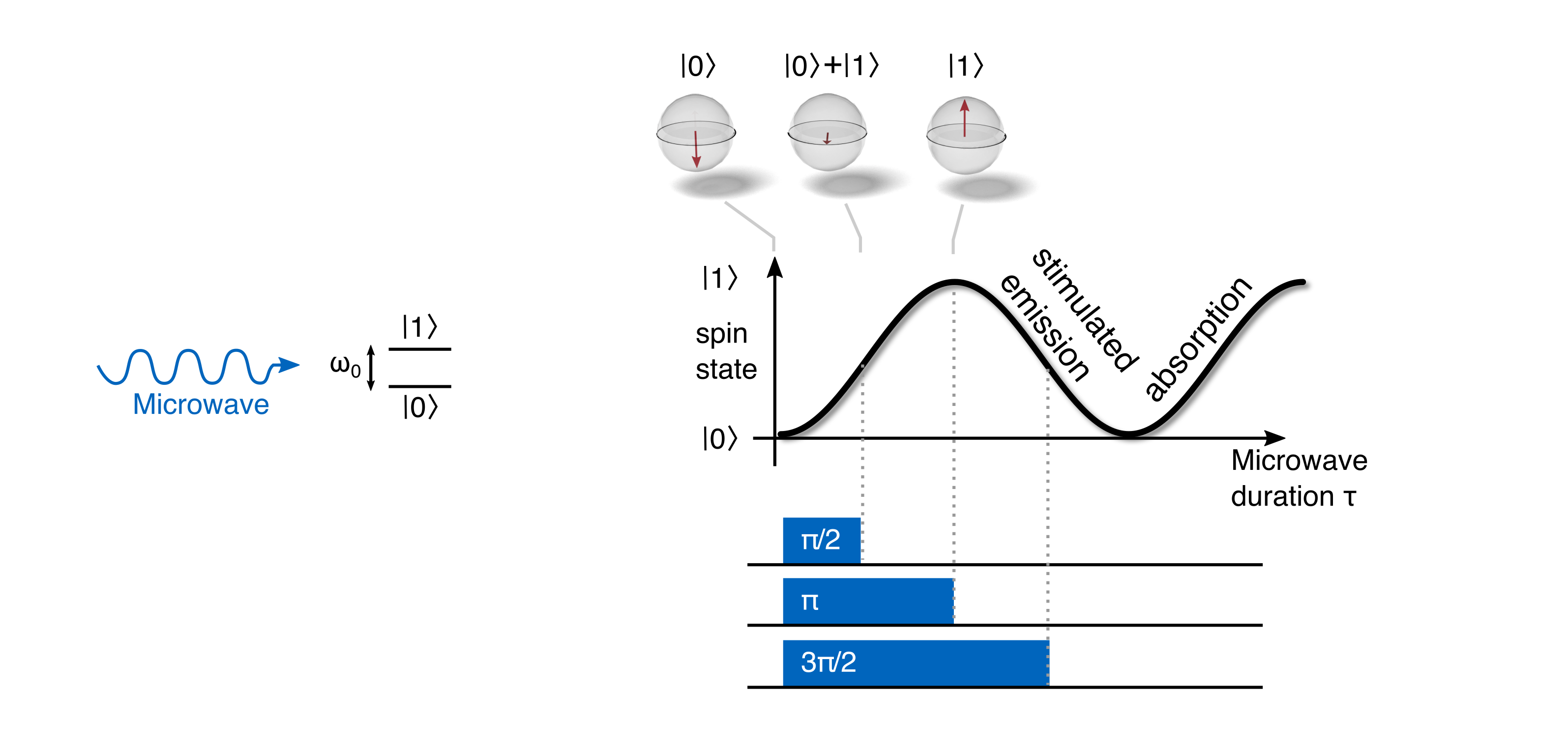}     
\caption{Rabi oscillations. A spin driven by a resonant microwave oscillates between states $\ket 0$ and $\ket 1$ by alternating phases of absorption and stimulated emission. A pulsed microwave drive can be applied for a full spin flip ("$\pi$-pulse") or can be stopped half-way ("$\pi/2$-pulse"), leaving the spin in a coherent superposition $(\ket 0 + \ket 1)/\sqrt 2$. }
\label{fig:rabi}
\end{figure}
\begin{figure}
\includegraphics{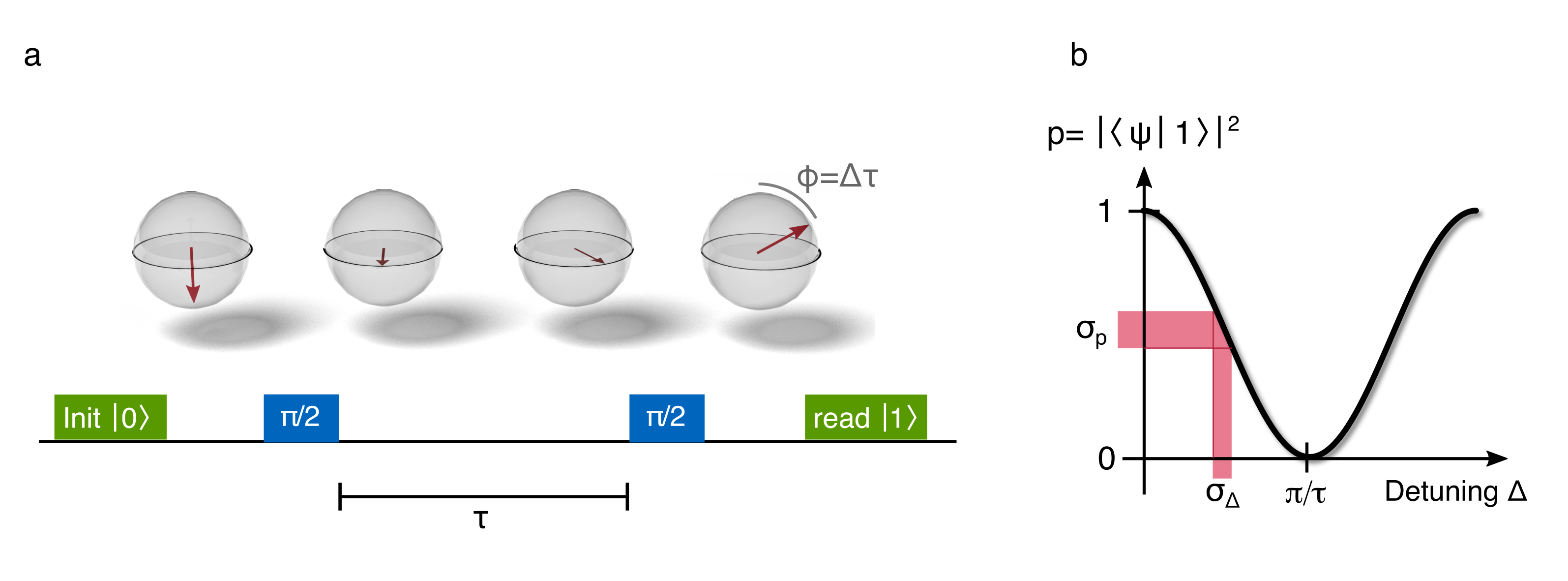}     
\caption{The Ramsey protocol. a) An initial $\pi/2$-pulse prepares a coherent superposition, which undergoes free evolution to pick up a phase $\phi=\Delta \tau$. This phase is converted into spin population by a second $\pi/2$-pulse. b) Sensitivity of Ramsey spectroscopy. For a fixed evolution time $\tau$, the spin state is an oscillatory function of the shift $\Delta$. Its state can be measured with uncertainty $\sigma_p$, which translates into an uncertainty $\sigma_\Delta$ on the measurement of $\Delta$. }
\label{fig:ramsey}
\end{figure}
The key idea behind these pulses can be understood by considering the evolution of a spin that is driven by a resonant microwave (Fig. \ref{fig:rabi}). A spin starting in state $\ket 0$ will absorb a photon from the microwave, flipping from $\ket 0$ to $\ket 1$. Somewhat counterintuitively, a spin starting from $\ket 1$ will flip back from $\ket 1$ to $\ket 0$, emitting a microwave photon by  stimulated emission. Under continuous driving, these processes will alternate, so that the spin rotates back and forth between $\ket 0$ and $\ket 1$, a process known as Rabi oscillation. This rotation revolves at an angular frequency $\Omega $, refered to as Rabi frequency, which is set by the power of the microwave drive. 
The effect of a microwave pulse depends on its duration. It can be set to perform half of an oscillation, flipping the spin from $\ket 0$ to $\ket 1$. Such a pulse is refered to as a $\pi$ pulse, since its duration $\tau$ is chosen such that $\Omega \tau = \pi$. Crucially however, the Rabi oscillation can also be stopped "half-way" between $\ket 0$ and $\ket 1$, leaving the spin in the coherent superposition $(\ket 0 + \ket 1)/\sqrt 2$. Since the duration of such a pulse is twice shorter (such that $\Omega \tau = \pi/2$), it is refered to as a $\pi/2$ pulse. \par
In a more rigorous manner, these effects can be derived from the Hamiltonian of atom-light interaction 
\begin{eqnarray*}
	\hat H &=& \frac\hbar 2 (\omega_0-\Delta)\hat \sigma_z + \hbar\Omega \cos(\omega_0 t) \hat\sigma_y \\
	&\stackrel{\text{rot. frame}}{=}& -\frac\hbar 2 \Delta\hat \sigma_z + \frac{\hbar\Omega}2 (1+e^{-2i\omega_0 t})\hat\sigma_y\\
	&\stackrel{\text{rot. wave approx.}}{\approx}&  - \frac\hbar 2 \Delta \hat \sigma_z + \frac{\hbar\Omega}2\hat \sigma_y
\end{eqnarray*}
Here the transformation into a rotating frame with frequency $\omega_0$ (step 1) not only renormalizes the qubit energy, it also introduces an oscillatory factor $e^{-i\omega_0 t}$ to all terms coupling states $\ket 0$ and $\ket 1$, in particular $\hat \sigma_y = i\ket 0 \bra 1 + H.C.$. In this process the two components  $\cos(\omega_0 t) = (e^{i\omega_0 t} + e^{-i\omega_0 t})/2$ of the cosine are transformed into a constant term $(1)$ and a rapidly oscillating term $(e^{2i\omega_0 t})$. This latter term can be neglected in a "rotating frame approximation" as long as $\Omega, \Delta \ll \omega_0$. \par
In this description, a microwave drive is equivalent to a rotation of the spin around axis $y$ of the Bloch sphere, induced by the term proportional to $\hat \sigma_y$. Typically, the drive is so strong that $\Omega \gg \Delta$, so that the term in $\hat \sigma_z$ can be neglected whenever the drive is applied. In these conditions, $\pi$ and $\pi/2$ pulses correspond to the propagators 


\begin{equation*}
\hat \pi = \exp\left[-i\frac{\Omega \tau_\pi}2 \hat \sigma_y\right] \qquad
\hat {\frac{\pi}2} = \exp\left[-i\frac{\Omega \tau_{\pi/2}}2 \hat \sigma_y\right]
\end{equation*}
with
$$
\hat \pi \ket 0 = \hat {\frac\pi 2} \cdot \hat {\frac\pi 2} \cdot \ket 0 = \hat {\frac{\pi} 2} \cdot (\ket 0 + \ket 1)/\sqrt 2 = \ket 1.
$$
With these ingredients we can understand the most fundamental quantum sensing protocol, the Ramsey sequence (Fig. \ref{fig:ramsey}). An initial $\pi/2$ pulse prepares the spin into the coherent superposition $(\ket 0 + \ket 1)/\sqrt 2$. It is subsequently subjected to free evolution under a spectral shift $\Delta$ for some waiting time $\tau$, picking up a phase $\phi=\Delta t$. A second $\pi/2$ pulse converts this phase to a population inversion of the spin state by  transforming the state from $(\ket 0 + e^{i\phi} \ket 1)/\sqrt 2$ to $i\sin{\frac\phi 2} \ket 0 + \cos{\frac\phi 2} \ket 1$. This population inversion  is measurable, since it affects the probability $\left|\bra 1 \psi\rangle\right|^2$ of finding the spin in state $\ket 1$ at the end of the sequence. This probability is an oscillatory function of $\phi$ 
$$
\left|\bra 1 \psi\rangle\right|^2 = \cos^2\left({\frac\phi 2}\right) = \frac 12 (1 + \cos {\phi}).
$$
If there is no precession during the free evolution $\tau$, the two $\pi/2$ pulses will add up to a $\pi$ pulse, leaving the spin in state $\ket 1$ at the end of the sequence. Precession during $\tau$ will be translated into a rotation between states $\ket 0$ and $\ket 1$. \par
Experimentally, $\phi $ can be varied both by varying $\tau$ and by varying $\Delta$. In both cases, the measurement result will trace an oscillatory function, refered to as Ramsey fringes in the time and frequency domain, respectively. For sensing applications, the latter implementation is more relevant. Here, the oscillating spin signal at the end of the experiment provides a ruler for spectral shifts $\Delta$, similar to the fringes of an optical interferometer providing a ruler for displacement. \par
Since quantum sensing experiments typically aim at detecting very small signals, the fringes are hardly ever recorded in their entirety. Instead, the sensor is operated at a fixed detuning $\Delta_0$, chosen such that the spin signal $p=\left|\bra 1 \psi\rangle\right|^2$ is maximally sensitive to small fluctuations in $\Delta$. This is typically the slope of a Ramsey fringe, and operation at this point can be ensured by applying a fixed detuning of $\Delta_0 = \pi/(2\tau)$ to the microwave drive. \par
The precision of such a sensing experiment is limited by the precision of spin readout. The measurement of $p=\left|\bra 1 \psi\rangle\right|^2$ will always have some experimental uncertainty $\sigma_p$, which translates into an uncertainty $\sigma_\Delta$ on the measurement of $\Delta$ by Gauss' law of error propagation.
$$
\sigma_\Delta = -\sigma_p \cdot \frac{d\Delta}{dp}\Huge|_{\text{steepest slope}} = \sigma_p \cdot \frac 2 \tau.
$$
If all technical fluctuations are eliminated, the fundamental limit of $\sigma_p$ is set by quantum projection noise. A single measurement of $p$ cannot access a floating point value between $0$ and $1$. Instead, spin readout will project the spin into either $\ket 0$ or $\ket 1$, providing one quantized bit of information ($p_\text{measured} = 0$ or $p_\text{measured} = 1$). Since we are operating at the slope, where $p\approx 0.5$, this measurement is always wrong, with an error of $$\sigma_p = \sqrt{\langle{p_\text{measured}^2}\rangle - \langle p_\text{measured}\rangle ^2} = \sqrt{1/2 - 1/4} = 1/2,$$ 
$\langle\cdot\rangle$ denoting the expectation value. As we average the results of $M$ repeated measurements on $N$ sensors operating in parallel, this error diminishes according to 
$$
\sigma_p = \frac 1 {2\sqrt {MN}}
$$
so that the sensor achieves a precision of
$$
\sigma_\Delta = \frac 1 {\tau \sqrt {MN}}.
$$
This result could have been obtained by common sense: a single measurement of duration $\tau$ is able to measure a frequency with a Fourier-limited resolution $1/\tau$, and averaging the result over $M\cdot N$ uncorrelated measurements will boost precision by a factor of $1/\sqrt {MN}$.\par
For simplicity, we have assumed single-shot readout, i.e. that the spin state can be measured in a single experimental repetition. In experiments, this is frequently impossible, for instance because of low detection efficiency of photodetectors or because the readout signal differs by less than 100\% between states $\ket 0$ and $\ket 1$. In this case, the number of repetitions $M$ has to be replaced by $M/M_0$, where $M_0$ denotes the number of measurements required to measure the spin state once.  
\subsection{Decoherence and the fundamental limit to sensitivity}
The above expression for $\sigma_\Delta$ is not yet a useful figure of merit for sensor performance, because $\sigma_\Delta$  can always be boosted to infinite precision by averaging over a larger number of measurements $N$ or by increasing the slope of the fringes by increasing $\tau$. Obviously, there are practical limits to both of these tricks that need to be taken into account by a reasonable figure of merit. In practice this is achieved by comparing sensor performance in terms of "sensitivity", a corrected figure of merit which we will derive in the following paragraphs.\par
Regarding averaging, it is important to note that this concept can be applied to any sensor. A statement like "an NV qubit can measure fields with nT precision" is useless (although we have made it above), since even a simple compass could in principle be pushed to this level by averaging its signal over a long time. To discount for this effect, sensor sensitivity is normalized to acquisition time. Since $M$ measurements will take  a time of $T=M\cdot \tau$, spectral resolution scales as $\sigma_\Delta = 1/(\sqrt{\tau N T})$, and the precision in sensing a magnetic field is 
\begin{equation}
\label{eq:sigma_B_tau}
\sigma_B = \frac{\sigma_\Delta}{\gamma} = \frac{1}{\gamma \sqrt{\tau N T}} = \eta_B \frac 1 {\sqrt T}
\end{equation}
Here, the {\bf sensitivity} $\eta_B = 1/(\gamma \sqrt {\tau N})$ provides a figure of merit that discounts for averaging. A sensor that can reach a higher precision in a given amount of averaging time is rated a better sensor, quantified by a lower value of sensitivity. Similar formulas can be established for other quantities ($E,T,\dots$) that can be measured by spectroscopy of some suitable spectral shift. The units of these figures are 
$$
\left[\eta_B\right] = \frac {\text{T}} {\sqrt \text{Hz}}, \qquad \left[\eta_E\right] = \frac {\text{V/cm}} {\sqrt \text{Hz}}, \qquad \left[\eta_T\right] = \frac {\text{K}} {\sqrt \text{Hz}},\qquad \dots
$$
The slightly un-intuitive unit $1/\sqrt{\text{Hz}}$ is omnipresent whenever sensitivities are involved. It is best understood as "sensor precision obtained after 1 second of averaging". \par
Still, sensitivity as defined in eq. \ref{eq:sigma_B_tau} is not a meaningful figure of merit, since it can be pushed to arbitrary limits by increasing $\tau$. The fundamental limit to this parameter is set by a quantum-mechanical effect known as decoherence, which limits $\tau$ to values less than the qubit's "coherence time" $T_2^*$. Decoherence refers to the fact that a quantum superposition $(\ket 0 + e^{i\phi} \ket 1)/\sqrt 2$ does not persist forever. It is a fragile state, which decays into a "classical mixture", where the qubit is in state $\ket 0 $ or $\ket 1$ with 50/50 probability. This state is different from the coherent superposition that the qubit started from, where the spin in a way lives simultaneously in both $\ket 0$ and $\ket 1$. The difference can be quantified by the density matrix 
$$
\hat \rho = \sum_i p_i \ket{\psi_i}\bra{\psi_i}
$$
where $p_i$ denotes the (classical) probability of the qubit to be prepared in a quantum state $\psi_i$ chosen from some set of states indexed by $i$. The density matrix $\hat \rho$ can capture the difference between "simultaneously $\ket 0$ and $\ket 1$" and "either $\ket 0$ or $\ket 1$ with 50/50 probability", since 
\begin{eqnarray*}
\hat\rho\left[(\ket 0 + e^{i\phi} \ket 1)/\sqrt 2\right] &=& \frac 12 
\left(\begin{matrix} 1 & e^{i\phi}\\ e^{-i\phi} & 1\end{matrix}\right)\\
\hat\rho_{\text{50/50}} &=& \frac 12 (\ket 0 \bra 0 + \ket 1 \bra 1) = 
\frac 12 
\left(\begin{matrix} 1 & 0\\ 0 & 1\end{matrix}\right)
\end{eqnarray*}
\begin{figure}
\includegraphics{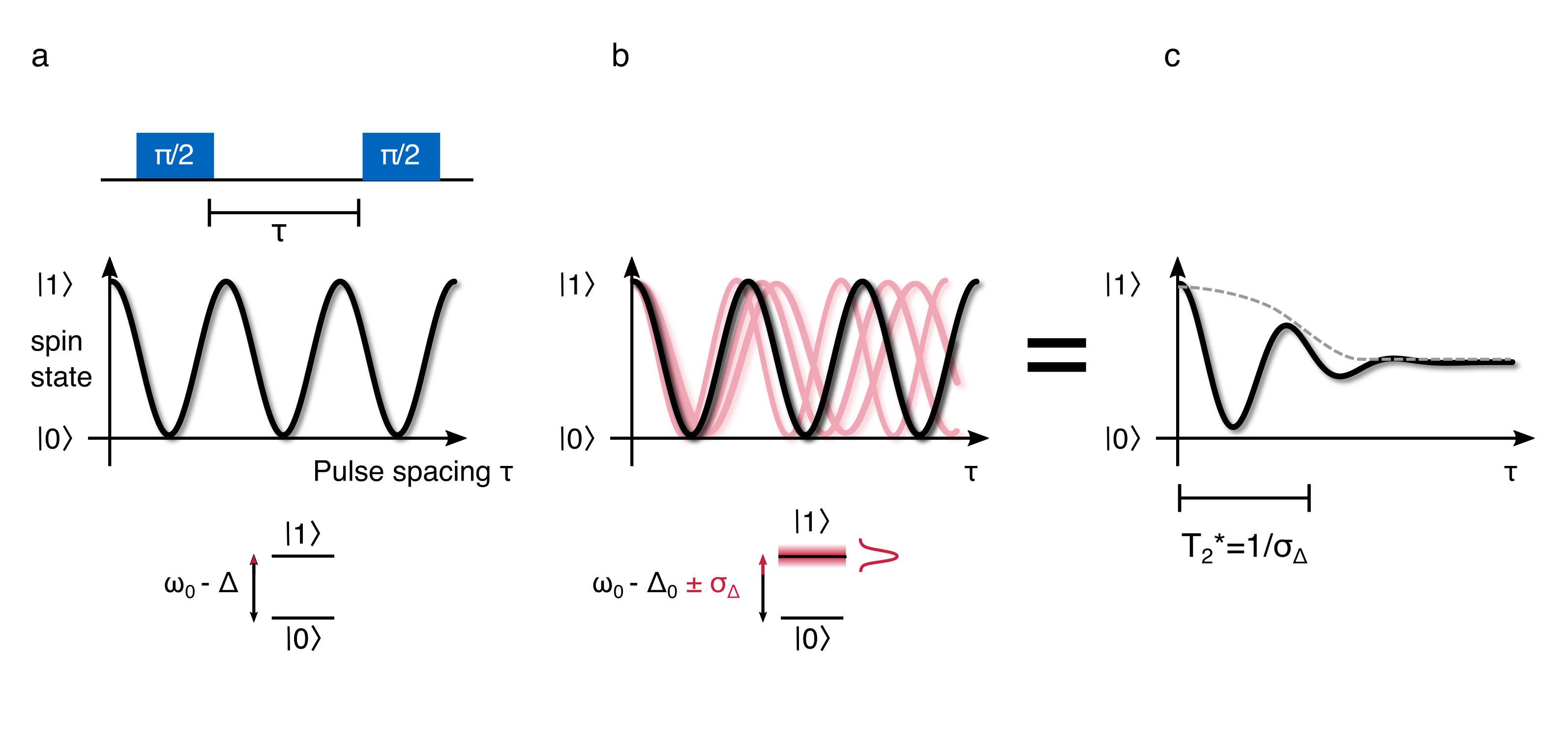}     
\caption{Decoherence. A random static background shift $\sigma_\Delta$ blurs the Ramsey fringes, inducing a decay of spin contrast over a timescale $T_2^* = 1/\sigma_\Delta$. }
\label{fig:decoherence}
\end{figure}
We see that quantum coherence is quantified by the off-diagonal elements of $\hat \rho$. To see why quantum states have a tendency to decay from a coherent state into classical mixtures, we consider a qubit that has some intrinsic uncertainty $\sigma_\Delta$ on its transition frequency $\Delta$ (Fig. \ref{fig:decoherence} b). This uncertainty could arise from a fluctuating magnetic background field, for instance the stray field from nuclear spins in the carbon lattice in the case of an NV center. We assume the background field to take a random value in each experimental repetition, so that the detuning $\Delta$ is normally distributed with a probability density
$$p(\Delta) = \frac 1 {\sqrt{2\pi\sigma_\Delta^2}} e^{-\frac{(\Delta-\Delta_0)^2}{2\sigma_\Delta^2}}$$
Each random value of $\Delta$ will lead to a different state $(\ket 0 + e^{i\Delta \tau} \ket 1)$ at the end of the free evolution, so that the spin state before the second $\pi/2$ pulse is described as the statistical mixture
$$
\hat\rho = \int d\Delta\, p(\Delta)\left(\begin{matrix} 1 & e^{i \Delta \tau}\\
e^{-i \Delta \tau} & 1 \end{matrix}\right) \stackrel{\text{Gaussian integral}}{=} \left(\begin{matrix} 1 & e^{- \frac 12 \sigma_\Delta^2 \tau^2 }e^{i \Delta_0 \tau} \\
e^{- \frac 12 \sigma_\Delta^2 \tau^2}e^{-i \Delta_0 \tau}  & 1 \end{matrix}\right)
\label{eq:decoherence}
$$
The off-diagonal elements of this state decay as $e^{- \frac 12 \sigma_\Delta^2 \tau^2}$ for increasing evolution time $\tau$, with a time constant of $$1/\sigma_\Delta := T_2^*.$$
Since it is the off-diagonal terms that are converted into spin population by the second $\pi/2$ pulse, this decay manifests itself as a decay of the time-domain Ramsey fringes over a timescale of $T_2^*$ (Fig. \ref{fig:decoherence} c). This result can also be derived from a more classical argument: a randomly varying  detuning $\Delta$ will lead to randomly varying frequencies of the time-domain Ramsey oscillations (Fig. \ref{fig:decoherence} b). For small evolution times $\tau$, this does not have a visible effect, since the spin will roughly end in the same peak or valley in any experimental repetition. For large evolution times, the measurement result is increasingly randomized, with the spin ending randomly in a peak or a valley depending on the exact value of $\Delta$ in a specific experimental repetition. Averaging over all these results can only lead to one result: $p=0.5$, zero spin contrast, no fringes. \par
Once decoherence is considered, the choice of $\tau$ in a sensing experiment is subject to a compromise. Too short values will lower sensitivity, because the frequency-domain fringes become less steep. Too long values will lower sensitivity, because decoherence reduces fringe contrast and, concomittantly, their slope. An educated guess suggests that the optimum choice is $\tau\approx T_2^*$, long enough to build up a steep slope, but sufficiently short to avoid death from decoherence. This guess is correct, as can be verified by an explicit minimization of the sensitivity as a function of $\tau$. With this choice, the sensitivity of a quantum sensor reads
\[
\tag{1}
\eta_B = \frac 1{\gamma \sqrt{N T_2^*} }
\label{eq:equation_1}
\]
 which is known as the "standard quantum limit". This relation is so fundamental that it is refered to as "equation (1)" by some researchers in the field \cite{budker07} and in this review. \par
We conclude this chapter by a discussion of some of its more subtle implications 
\begin{itemize}
	\item It is a frequent misconception that a stronger coupling to the environment $\gamma$ will automatically yield a sensor with a better sensitivity, suggested by the fact that $\gamma$ appears in the denominator of eq. (\ref{eq:equation_1}). The fallacy is that a stronger coupling to the environment also impacts $T_2^* = \sigma_\Delta^{-1}$ by increasing the frequency noise $\sigma_\Delta$. A famous example of this subtlety is the use of entangled states of multiple qubits such as $(\ket {00} + e^{i\phi}\ket{11})/\sqrt 2$. These have long been believed to result in better sensitivity than a sensor built from independent qubits. Since their total magnetic moment grows linearly with $N$, an entangled sensor employing a single effective spin of $N$ entangled qubits would at first sight achieve a sensitivity of $1/(\gamma N \sqrt{T_2^*})$, a factor of $\sqrt N$ better than eq. (\ref{eq:equation_1}) \cite{wineland92, wineland94, giovannetti04}. Unfortunately, stronger coupling also increases the impact of background noise, so that increased decoherence shrinks $T_2^*$ by a factor of $N$, spoiling the improvement \cite{huelga97, giovannetti11}. 
	\item While we frequently speak of $T_2^*$ as an intrinsic constant of a specific qubit, we should think of it just as much as a constant of a qubit's environment, quantifying the level of intrinsic background noise. As a specific example, NV centers have orders of magnitude higher coherence times than quantum dots, and we might easily think of them as "better" qubits. Yet, both are based on the same quantum entity, a single electron spin, and the difference merely results from the fact that nuclear spins are a lot less abundant in diamond than in typical quantum dot materials like GaAs. As a consequence, the relation might turn upside down if quantum dots could be engineered in diamond, and the search for new spin qubits could result in disappointment if it is focused on the wrong materials. Most importantly, both qubits could become equally bad sensors when they have to be operated close to a sample with high intrinsic noise. 
\end{itemize}
\section{Sensing by dynamical decoupling - the hidden revolution}
\begin{figure}
\includegraphics{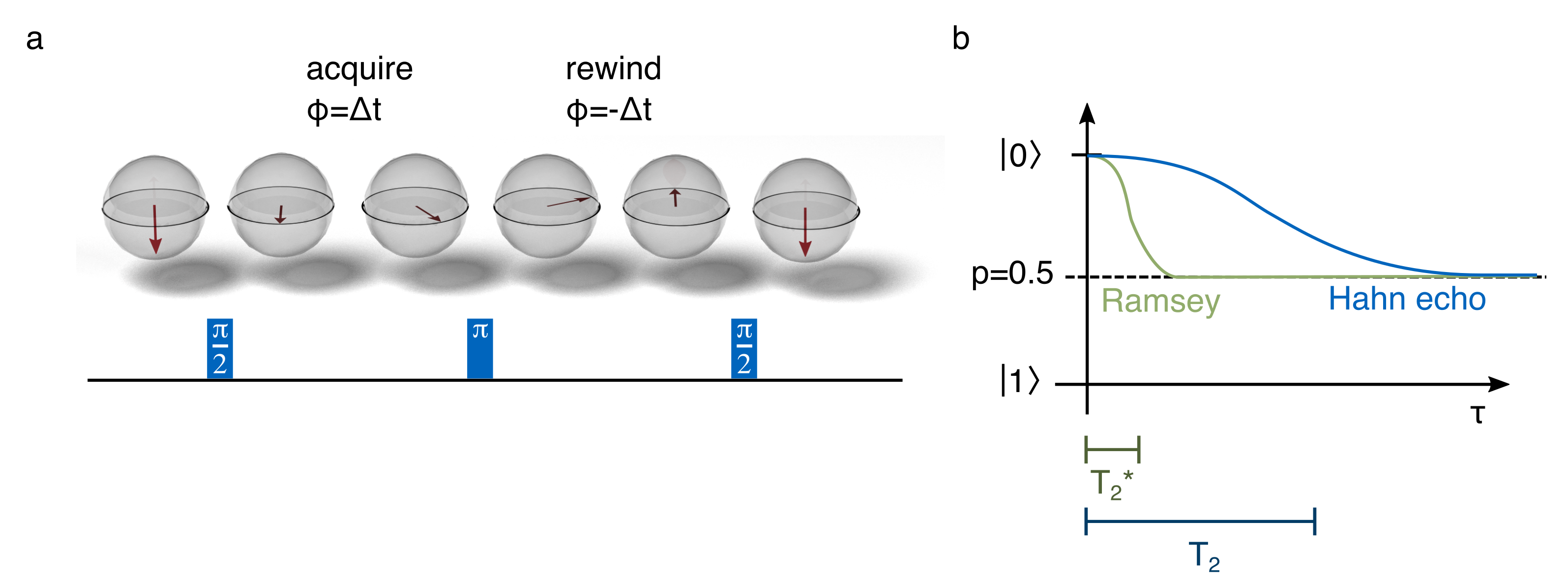}     
\caption{Hahn echo. a) A $\pi$-pulse in the middle of the free evolution time mirrors the spin to the back side of the equator, so that a phase acquired from a static shift $\Delta$ is exactly rewound in the second half of the sequence. b) Coherence persists for a longer time $T_2$ under the Hahn echo sequence.}
\label{fig:hahn_echo}
\end{figure}
Despite its fundamental nature, equation (\ref{eq:equation_1}) can be broken. The key idea here is that $T_2^*$ can be extended to longer timescales if a tool is found to suppress buildup of a random phase $\sigma_\Delta \tau$ in presence of a random spectral shift $\sigma_\Delta$. Such a tool is provided by quantum control protocols, more complicated manipulations of the spin than a mere creation and detection of coherence. The first protocol of this kind, known as "spin echo" or "Hahn echo", was developed by Erwin Hahn in 1950~\cite{hahn50}. It is presented in Fig. \ref{fig:hahn_echo}. It differs from standard Ramsey spectroscopy by an additional $\pi$-pulse, introduced in the middle of the free evolution time. This pulse mirrors the spin state to the other side of the equator, so that a phase built up by a constant detuning $\Delta$ in the first half of evolution is exactly rewound in the second half. By this mechanism, the effect of a constant $\Delta$ is canceled and decoherence is suppressed, so that $\tau$ can be chosen much longer than $T_2^*$. In practice there is a new limit to $\tau$, refered to as $T_2$, arising from the fact that $\Delta$ is never purely static and that the spin phase remains sensitive to fluctuations that vary over the timescale of the Hahn echo sequence. This timescale is situated between $T_2^*$ and $T_1$, the spin-flip lifetime of an incoherent spin state like $\ket 1$. \par
\begin{figure}
\includegraphics{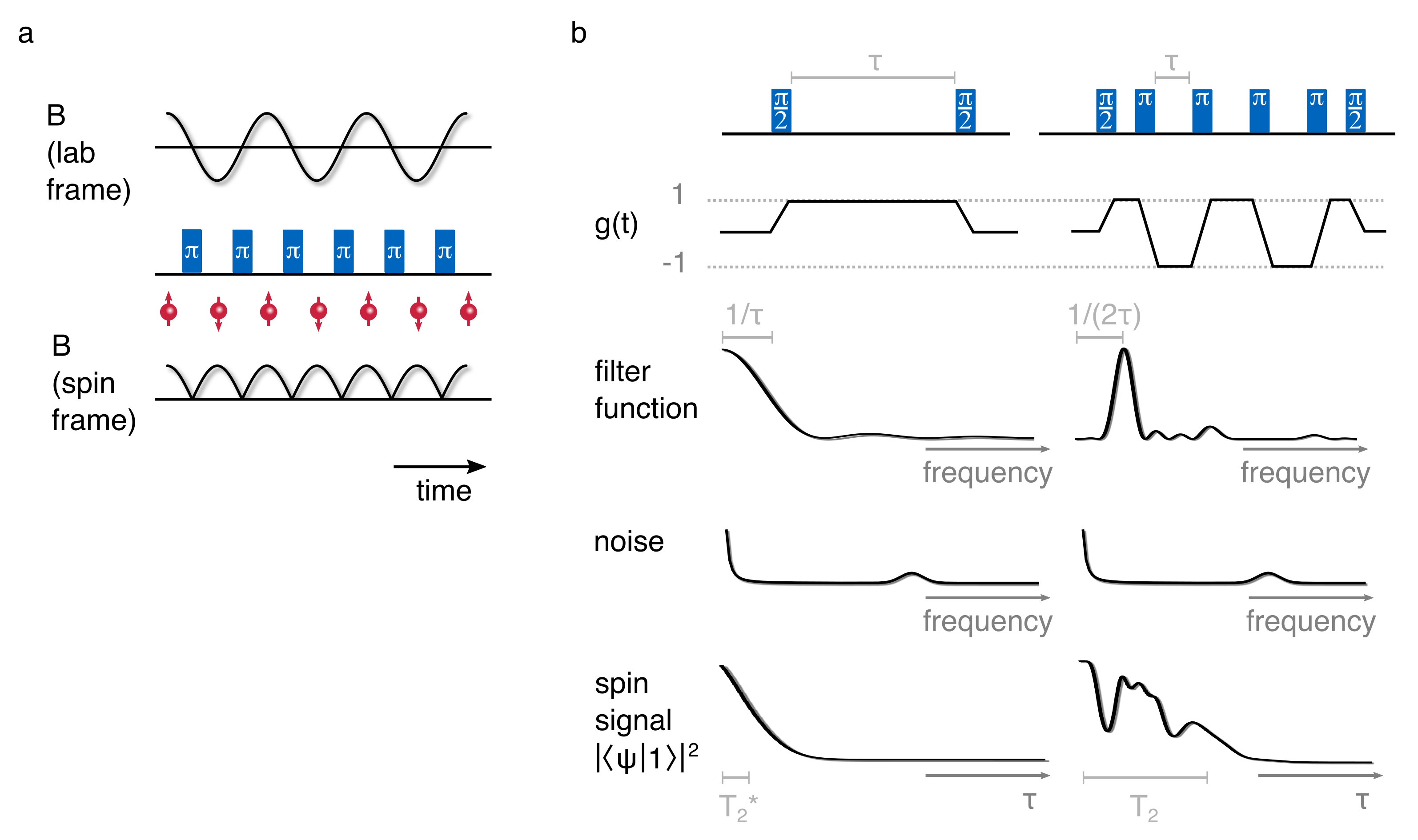}     
\caption{Dynamical decoupling. a) a train of $\pi$-pulses acts as a lock-in grating, effectively inverting the sign of the signal in the flipping reference frame of the sensor spin. b) The spectral sensitivity of a spin under dynamical decoupling is modeled by a "filter function", the Fourier transform of a time-domain sensitivity function $g(t)$. For a $\tau$-periodic grating (right), the filter function is peaked at a frequency $1/(2\tau)$, so that the spin coherence decay traces an inverted copy of the noise spectrum, turning the spin into a spectrum analyzer.}
\label{fig:dynamical_decoupling}
\end{figure}
At first sight the Hahn echo protocol appears to be useless for sensing, since it cancels a constant detuning $\Delta$. While this mitigates decoherence, it comes at the price of canceling a static signal, too. It turns out, however, that a qubit subjected to a Hahn echo sequence remains sensitive to signals oscillating at a nonzero frequency and that this sensitivity can actually be tuned to specific signal frequencies in a very selective manner~\cite{alvarez11, delange11, kotler11, delange10, du09, biercuk09, viola98, uhrig07}. The key idea, refered to as "quantum lock-in detection" or "dynamical decoupling" is presented in \ref{fig:dynamical_decoupling} a. We consider a qubit that is exposed to an oscillating signal at a frequency $\nu$. This signal could not be detected in a Ramsey sequence, since positive and negative half-cycles would imprint opposite phases on the spin and cancel over the free evolution time. We can, however, effectively rectify the signal by applying periodic $\pi$ pulses to the spin, whose spacing $\tau$ is synchronized to the signal by the choice $2\tau = \nu^{-1}$. Since each pulse induces a flip of the sensor spin, it effectively inverts the sign of the signal in the flipping reference frame of the sensor. The effect of this scheme is similar to classical lock-in-detection, where the signal would be inverted by a mixer or a similar device. Signals at a frequency $1/(2\tau)$ are rectified to DC (zero frequency) while static shifts will be converted to some nonzero frequency so that their effect cancels over time. \par
To employ this scheme in a sensing sequence, the lock-in grating of periodic $\pi$-pulses is embedded into two $\pi/2$-pulses to create and read a quantum phase (\ref{fig:dynamical_decoupling} b). The $\pi$-pulse grating tunes sensitivity to a specific frequency, which simultaneously suppresses decoherence from static shifts and sensitizes the quantum phase to specific AC signals. With dynamical decoupling, a qubit becomes a quantum spectrum analyzer. Ramsey spectroscopy and Hahn echo can be regarded the two simplest incarnations of this concept, tuning sensitivity to DC and $1/(2\tau)$ respectively, but a whole multitude of novel experiments can be conceived by playing with the placement of $\pi$ pulses. \par
The spectral sensitivity of an arbitrary decoupling sequence can be computed by a formalism known as the "sensitivity function" or "filter function"~\cite{cywinski08, alvarez11}. In this formalism the phase that a qubit picks up over a dynamical decoupling sequence is written as 
$$
\phi = \int dt\; g(t) \Delta (t)
\label{eq:sensitivity_function}
$$
where $g(t)$ is a sensitivity function, modeling the effective inversion of the signal by the $\pi$ pulses. It can be constructed from a few simple rules. The initial $\pi/2$  pulse initializes $g$ with $g(t) = 1$, since it creates a coherent superposition whose phase is sensitive to $\Delta$. $g$ changes sign with every $\pi$-pulse to model the effective inversion of the signal in the flipping frame of the spin. Finally, $g(t)=0$ before and after the initial and final $\pi/2$-pulses, since there is no quantum coherence beyond these points. In an alternative interpretation, $g$ can be understood as the "Platonic ideal" of a signal that will impart a maximum phase $\phi$ on the qubit. 
\par
In the most general case, a signal does not have a constant intensity and phase, so that the quantum phase $\phi$ is a random variable with a variance $\langle\phi^2\rangle$. The spin signal  $|\bra \psi 1\rangle|^2$ at the end of the sequence can be computed from eq. (\ref{eq:decoherence})
$$
|\bra \psi 1\rangle|^2 = \frac 12 \left(1 + e^{-\langle\phi^2\rangle / 2}\right).
\label{eq:decoherence_phi}
$$
In absence of a signal $(\langle\phi^2\rangle=0)$ coherence will be preserved, so that the final $\pi/2$-pulse flips the spin into $\ket 1$. A nonzero signal will induce decoherence $(\langle\phi^2\rangle>0)$, reducing the spin signal. We now rewrite eq. (\ref{eq:sensitivity_function}), expressing the $g$ and $\Delta$ involved by their Fourier coefficients $g_n, \Delta_n$
\begin{eqnarray*}
\phi &=& \int_0^T dt \; g(t) \Delta(t) \\
&\stackrel{f(t)=\frac 1T \sum_{-\infty}^\infty f_n e^{i2\pi n t/T}}{=}&
\frac 1 {T^2} \sum_{n, n'} g_n \Delta_{n'}\int_0^T dt \; e^{i 2\pi n t/T}e^{i 2\pi n' t/T} \\
&\stackrel{\int\dots = T\delta_n^{n'}}{=}&
\frac 1 {T} \sum_n g_n \Delta_{-n}
\end{eqnarray*}
Here, $T$ denotes the full length of the free evolution from the initial to the final $\pi/2$ pulse. The variance $\langle \phi^2\rangle$ becomes thus
\begin{eqnarray*}
\langle\phi^2\rangle &=& 
\frac 1 {T^2} \left< \sum_{n, n'} g_n^*\Delta_{-n}^* g_{n'} \Delta_{-n'}\right> \\
&\stackrel{\langle \Delta_n \Delta_{n'}^*\rangle = 0}{=}&
\frac 1 {T^2}\sum_n \left|g_n\right|^2 \left|\Delta_n\right|^2\\
&=& \sum_n S_g (\nu_n) S_\Delta (\nu_n)
\end{eqnarray*}
Here, the relation $\langle \Delta_n \Delta_{n'}^*\rangle = 0$ holds, since different Fourier components of noise are uncorrelated. $S_f(\nu_n) = |f_n|^2/T$ denotes the power spectral density of a quantity $f$. This final relation underlines the above statement that a dynamical decoupling sequence sensitizes a qubit to noise at specific frequencies, described by the Fourier transform $S_g(\nu_n)$ of the sensitivity function $g(t)$. \par
Some practical examples of this concept are presented in Fig. \ref{fig:dynamical_decoupling} b. Ramsey spectroscopy (left) has a constantly positive  sensitivity function in the time domain, resulting in a filter function with DC sensitivity in the frequency domain. It is insensitive to signals with a frequency $\nu \gg \tau^{-1}$, which average to zero over the free evolution. Accordingly, the frequency-domain filter function is a lowpass filter with bandwidth $1/\tau$. Since noise is mostly stronger at low frequencies, the spin signal decays on a fast timescale $T_2^*$ under Ramsey spectroscopy. \\
Under dynamical decoupling (specifically: the Carr-Purcell-Meiboom-Gill sequence, right), the time-domain sensitivity function periodically alternates between $+1$ and $-1$. The Fourier transform of this signal is peaked at a frequency $1/(2\tau)$, and vanishes at DC, modeling the fact that some AC signals are rectified while DC signals are suppressed as discussed above. The sequence can be used as a quantum spectrum analyzer by varying $\tau$. This will displace the peak of maximum sensitivity across a wide range of frequencies. The spin signal traces an inverted copy of the noise spectrum, with strong decoherence occuring whenever the sensitivity peak coincides with a peak of noise. Since the sequence equally supresses decoherence from DC background fields, coherence time under CPMG rises to $T_2$. By virtue of this second aspect, sensitivity for detection of AC signals is higher than the sensitivity obtained in a Ramsey sequence, and equation (\ref{eq:equation_1}) is modified to 
$$
\eta_B = \frac 1 {\gamma \sqrt{T_2 N}}.
$$
\par
\begin{table}[h]
  \caption{Sensitivity of magnetic field sensors. Grey shaded numbers denote extrapolations from published values. Numbers differ from the shot noise limit  of eq. ($\ref{eq:equation_1}$) because the efficiency of spin readout is not unity in most experiments. SQUID: Superconducting Quantum Inferference Device.}
  \label{tab:pricesI}
  \begin{tabular}{rp{1.2cm}p{1.2cm}p{1.5cm}p{1.5cm}p{1.5cm}p{1cm}l}
    \hline
            & single NV  & single NV ($^{12}$C) & NV\par ensemble & NV\par ensemble\par theory    & Smartphone\par Hall\par sensor& SQUID \\
    \hline
	  $N$ & $1$& $1$ & $10^{11}$ ($100\mu$m)$^3$ & $10^{16}$ ($1$cm$^3$)& & &\\
      $T_2^*$ & $1\mu$s & $228\mu$s & $30\mu$s & - & &\\
      $T_2$ &  $300\mu$s & $2$ms & $50\mu$s & $300\mu$s & &   \\
      $\eta_{\text{DC}}$  & $1\mu$T$/\sqrt{\text{Hz}}$ & \textcolor{grey}{$20$nT$/\sqrt{\text{Hz}}$} &  \textcolor{grey}{$1$pT$/\sqrt{\text{Hz}}$} & - & $30$nT$/\sqrt{\text{Hz}}$ & $1$fT/$\sqrt{\text{Hz}}$\\
      $\eta_{\text{AC}}$  & $20$nT/$\sqrt{\text{Hz}}$ & $4$nT$/\sqrt{\text{Hz}}$ & $1$pT$/\sqrt{\text{Hz}}$ & $250$aT$/\sqrt{\text{Hz}}$ &  &  $20$aT$/\sqrt{\text{Hz}}$\\
	  Refs. & \cite{taylor08} & \cite{zhao12, balasubramanian09} & \cite{wolf15, bauch18, clevenson14}& \cite{taylor08}& \cite{akm10} & \cite{kominis03, simmonds79}\\
    \hline
  \end{tabular}
\end{table}

The effect of dynamical decoupling can be profound. For NV centers, $T_2$ can be four orders of magnitude longer than $T_2^*$ ($10$ms~\cite{farfurnik15} rather than $1\mu$s), so that sensitivity improves by two orders of magnitude whenever dynamical decoupling can be employed. Moreover, the ability to perform spectroscopy of a time-dependent magnetic signal is a powerful benefit. \par
Arguably the most prominent example for the power of dynamical decoupling is the detection of magnetic resonance signals from nanoscale samples, first achieved in 2013~\cite{mamin13, staudacher13}. Here, a single stationary NV center, embedded few nanometers beneath the surface of a bulk diamond, is employed as a quantum spectrometer to detect magnetic noise from samples on the diamond surface. This magnetic noise is mostly dominated by stochastic fluctuations of nuclear spins, peaked at the characteristic Larmor frequency of the molecules on the surface. Dynamical decoupling has pushed the sensitivity of a single NV center into a range where detection of this spin noise is possible. Both nuclear spins~\cite{lovchinsky16} and electron spin signals~\cite{shi15} have been detected from single biomolecules. The door seems open to transform magnetic resonance spectroscopy, previously limited to mm-scale samples or complicated low-temperature setups~\cite{degen09}, into a single-molecule technique. 

\section{Outlook - prospects and hopes after one decade}
Despite the impressive progress over the past decade, the achievements of nanoscale sensing by spin qubits have covered only a fraction of its full potential. Several important goals remain open and great further breakthroughs are likely to emerge:
\begin{itemize}
\item {\bf Scanning probe magnetometry} is still mostly based on measurements of strong DC fields, using the Ramsey sequence or resonant spectroscopy. Dynamical decoupling is hardly ever employed, so that orders of magnitude in sensitivity remain to be harvested. The reason for this surprising fact is that spin coherence is strongly degraded in nanodiamonds and nanofabricated diamond tips. Clearly, this  problem is not fundamental, and will be overcome by improved fabrication, more robust color centers or novel concepts for scanning probe positioning~\cite{ernst18}. With better sensitivity, novel samples will shift into reach. Likely targets are scanning-probe imaging of nuclear spins - so far limited to some proof-of-principle experiments in "best-case" conditions~\cite{rugar15,haeberle15}, imaging of current patterns in solid-state samples~\cite{chang17}, or spectroscopy of magnetic fluctuations as a tabletop complement to neutron scattering experiments in materials science~\cite{kolkowitz15, ariyaratne18}. 
\item {\bf Single-molecule magnetic resonance} is so far limited to a mere spectroscopy technique and suffers from very slow data acquisition. Much of the power of its bulk counterpart derives from imaging, most evident in the beautiful images provided by clinical scanners. Translating this technique to a three-dimensional imaging method for single molecules is technically challenging, but seems to be doable in the years to come. Improved strategies for signal acquisition like Fourier processing~\cite{boss17, schmitt17, arai15} and single-shot readout will be crucial to keep acquisition times at a manageable level. 
\item {\bf Sensing using spins in nanoparticles} has led to proof-of-principle measurements of temperature inside living cells~\cite{kucsko13, alkahtani17}. Progress is being made towards monitoring of chemical reactions~\cite{rendler17}. Also, fluorescent defect centers in nanoparticles have been used for superresolution microscopy~\cite{rittweger09, pfender14}. This application does not make use of any spin properties and merely exploits the fact that the fluorescence of color centers is photostable. All of these techniques have not found widespread adoptance yet, presumably because directed attachment to specific parts of a cell is difficult, owing to the complex surface chemistry of nanoparticles.
\item While much work has been done using single spins as detectors for nanoscale signals, all techniques discussed in this review can be scaled to {\bf sensing on the micron-scale}, using ensembles of spins as slightly larger detectors with better sensitivity. This line of research has led to magnetic microscopy of inclusions in meteorites~\cite{fu14} and bacteria~\cite{lesage13} and, very recently, to NMR spectroscopy on $\mu$m-scale samples~\cite{glenn18}. We might soon see commercial magnetic microscopes and microfluidic NMR detectors. With these detectors, magnetic resonance microscopy could ultimately  bridge all length scales between atomic resolution and the mm-scale resolution of clinical scanners. 
\item Finally, the progress of the past decades can be seen as a success of quantum control. In a first wave, manipulation of quantum states has been achieved in very clean artificial systems like trapped ions or ultracold atoms in vacuum. In a second wave, these achievements were extended to some selected solid state systems. A logical third challenge could be control of {\bf arbitrary spins in arbitrary samples}. Indeed, many streams of work are directed towards this goal. NV centers can not only be used to detect nuclear spins, but also to prepare them in a specific state by polarisation exchange. This could lead to a novel way for nuclear hyperpolarisation of arbitrary samples, a direction of research that is even starting to be pursued by industry. If nanoscale magnetic resonance could be pushed down to the level of single nuclei, solid state phases could be studied in real materials with atomic resolution, providing a third way between top-down studies on bulk samples and bottom-up replication of matter in ultracold atoms. Dynamical decoupling as a generic tool to tailor the spectral properties of materials could become the key to translate quantum effects like slow light to a much wider range of materials~\cite{joas17}. 
\end{itemize}


 

\bibliography{bibliography}{}
\bibliographystyle{plain}


\end{document}